\documentclass[prl,notitlepage,twocolumn]{revtex4}

\usepackage{amsmath}
\usepackage{amssymb}
\usepackage{bm}
\usepackage{graphicx}
\usepackage{times}
\usepackage{xcolor}
\usepackage{float}
\usepackage{gensymb}

\usepackage{pdfcomment}
\pdfcommentsetup{draft} 
\pdfcommentsetup{color={1.0 1.0 0.0}, open=true}

\usepackage{color}
\usepackage{ulem}
\usepackage{bm}
\usepackage{braket}
\usepackage{physics}
\newcommand{\ve}[1]{\mathbf{#1}}

\begin{document}

\title{High-Q localized states in finite arrays of subwavelength resonators}

\author{Danil~F.~Kornovan$^1$}
\author{Roman~S.~Savelev$^1$}
\author{Yuri~S.~Kivshar$^{1,2}$}
\author{Mihail~I.~Petrov$^1$}
\affiliation{$^1$Department of Physics and Engineering, ITMO University,  St.-Petersburg 197101, Russia}
\affiliation{$^2$Nonlinear Physics Center, Research School of Physics, Australian National University, Canberra ACT 2601, Australia}

\begin{abstract}
We introduce a novel physical mechanism for achieving giant quality factors ($Q$-factors) in finite-length periodic arrays of subwavelength optical resonators. The underlying physics is based on interference between the band-edge mode and another standing mode in the array, and the formation of spatially localized states with dramatically suppressed radiative losses. We demonstrate this concept for an array of $N$ dipoles with simultaneous cancellation of multipoles up to $N$-th order and the $Q$ factor growing as $Q \propto N^{\alpha}$, where $\alpha \gtrsim 6.88$. Based on this finding, we propose a realistic array of Mie-resonant nanoparticles ($N \lesssim 29$) with a dramatic enhancement of the Purcell factor (up to $\sim $3400) achieved by tuning of the array parameters.
\end{abstract}

\maketitle

Subwavelength trapping and guiding of light has attracted a great attention because it provides unique opportunities for miniaturization of the optical interconnect technology. Diverse implementations of subwavelength-engineered structures in integrated optics have been discussed for a design of integrated photonic platforms~\cite{Cheben2018} and for engineering the mode dispersion and waveguide anisotropy to achieve subwavelength propagation of slow light \cite{Figotin2011,Ding2020}.

For many years, plasmonics was considered as the only available platform for nanoscale optics, but the recently emerged novel field of Mie resonant metaphotonics provides practical alternatives for nanoscale localization of light by employing resonances in high-index dielectric nanoparticles and structures \cite{Koshelev2020_2,Won2019}. Future technologies underpinning high-performance optical communications, ultrafast computations and compact biosensing will rely on densely packed reconfigurable optical circuitry based on nanophotonics. Importantly, in the last few years we observe novel trends in achieving high values of the quality factor (Q-factor) in dielectric structures  \cite{Koshelev2019} for spatial and temporal control of light by employing multipolar resonances and their interference.

High-index dielectric nanoantennas supporting multipolar Mie resonances represent a novel class of building blocks of metamaterials for generating~\cite{Rutckaia2017a,Tiguntseva2020,Mylnikov2020}, manipulating~\cite{Kamali2018}, and modulating~\cite{Ding2020} light. By combing both electric and magnetic multipolar modes, one can not only modify far-field radiation patterns but also localize the electromagnetic energy in open resonators by employing the physics of bound states in the continuum~\cite{Hsu2016} to achieve destructive interference of two (or more) leaky modes \cite{Rybin2017,Koshelev2020}.

\begin{figure}[t]
\centering
\includegraphics[width=0.43\textwidth]{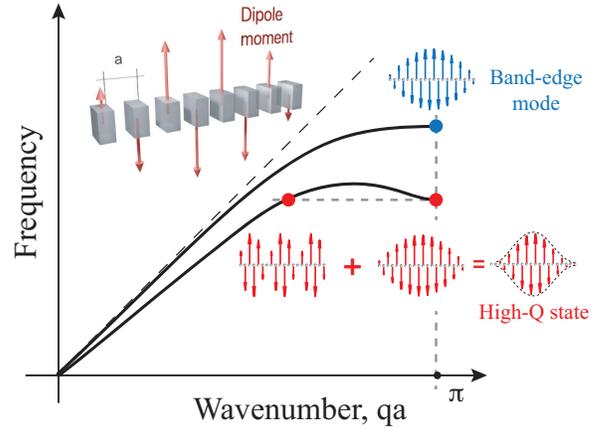}
\caption{An inflection point in the dispersion curve of an infinite one-dimensional dipole array of Mie-resonant particles (an upper inset) enables interference between the staggered  band-edge  mode and another standing mode of a finite-extent array, thus suppressing radiative losses with the formation of a high-$Q$ localized state.  }
\label{fig:1}
\end{figure}

The further increase of the $Q$-factor of resonant systems can be achieved by combining several subwavelength resonators which can provide additional degree of freedom in suppressing radiative losses. One of the simplest geometries is a finite-extend array of coupled dielectric resonant structures which supports a set of standing waves \cite{Savelev2014,Krasnok2016,Bakker2017}, as shown schematically in the insert of Fig.~\ref{fig:1}.  Importantly, at the frequencies close to the band edge the phases of the fields in the neighbour resonators become close to $\pi$, which leads to destructive interference between the field scattered by individual resonators and consequently to suppressed radiation from the whole structure.  Such band-edge (BE) localized states have been explored in applications to photonic crystals~\cite{Othman2016}, and they have already been  utilized for light emitting~\cite{Agrawal1986,Dowling1994,Kim2011,Hoang2020,Rutckaia2020} and sensing~\cite{Cusano2016,Yang2020} applications.

\begin{figure*}[htbp]
\centering
	\includegraphics[width=0.8\textwidth]{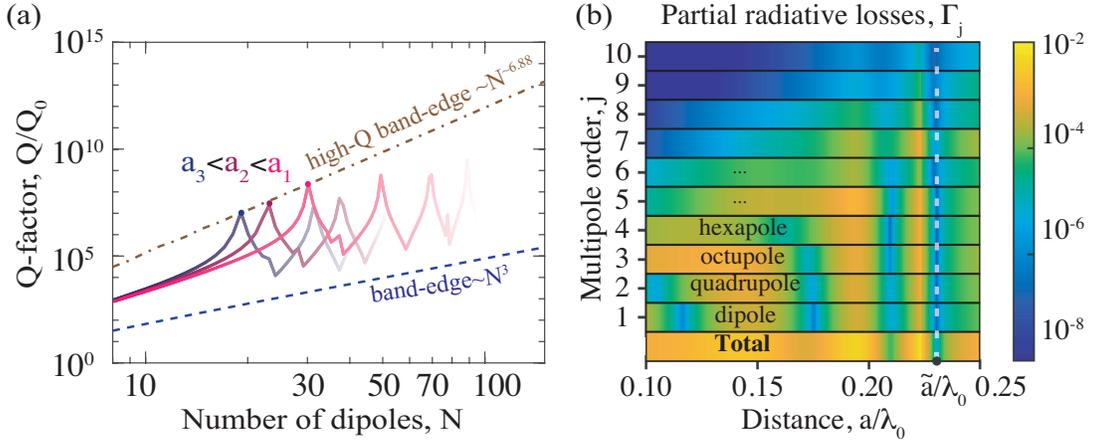}
\caption{Radiation suppression in arrays of $N$ dipoles. (a) The $Q$-factor of the resonant modes with different separation distances $a$. Blue dashed line corresponds to the band-edge mode with $a_{BE}/\lambda_0=0.35$, while three solid lines show the $Q$-factor dependence for three values of separation distance $a/\lambda_0=0.238, 0.239,\text{and } 0.240$ with maximal $Q$-factors at $N=19, 22, \text{and } 29$, respectively. Brown dash-dot line denotes the asymptotic dependence of the maximal $Q$-factors.  (b) The total   and partial radiative losses correspondent to different multipoles orders $j$ for the high-$Q$ state as a function of the spacing $a$ for $N=10$.  }
\label{fig:2}
\end{figure*}

Here, we introduce and explore {\it a novel physical mechanism} for achieving giant $Q$-factors in dielectric structures of subwavelength optical resonators. We consider an array of resonators and assume that, due to their long-range interaction, the dispersion curve can have a maximum, as schematically shown in Fig.~\ref{fig:1}. Then, for the frequency selected at the band edge, a staggered mode can couple resonantly with another standing mode supported by the array, so that we observe the formation of a hybrid localized state with dramatically suppressed radiative losses due to several interference mechanisms.  First, we demonstrate this general concept for an array of  $N$ dipoles that can support the modes with the $Q$-factor growing as $ Q\propto N^{\alpha}$, where $\alpha \sim 6.88$. Second, we show the formation of a high-$Q$ state accompanied by simultaneous resonant suppression  of high-order radiation multipoles.  Finally, we consider a realistic structure based on silicon-on-insulator resonator arrays and demonstrate a dramatic Purcell enhancement up to $\sim$ 3400 achieved by tuning parameters of the array.

{\it Dipole arrays.}
Our starting point is a simple model describing a one-dimensional periodic array of resonant dipoles with polarizability $\alpha(\omega)$ and lattice spacing $a$. The electromagnetic coupling between the dipoles leads to a self-consistent system of equations that describes collective dipolar oscillations supported by the structure [see Eq.~(S1) in Supplemental Materials~\cite{SuppMat}]. Solution of these equations provides us with a set of complex eigenfrequencies $\omega=\omega'-i\gamma$, where $\gamma$ determines the decay rate of the corresponding mode and the ratio $\omega'/(2\gamma)$ is known as the $Q$ factor.

The $Q$ factor calculated within the quasi-resonant approximation (see Supplemental Materials~\cite{SuppMat}), normalized by the $Q$ factor of an individual dipole $Q_0$, is shown in Fig.~\ref{fig:2}(a) as a function of the number of dipoles $N$ and for different values of the period $a$.
For the period $a_{BE}=0.35\lambda_0$, where $\lambda_0$ is the free space wavelength, we observe a typical growth $Q\propto N^3$ inherent to the regular band-edge states. Such dependence is dictated by the type of the dispersion near the edge of the Brillouine zone~\cite{Zhang2020}, which in general case is quadratic. Decrease of the period leads to modification of the dispersion curve and, for a specific value of the period (see analytical expression in Supplemental Materials \cite{SuppMat}) the group velocity dispersion vanishes and the dispersion curve becomes quartic. Such a type of dispersion is known to exist in different systems, e.g. in anisotropic crystals~\cite{Figotin2005} or coupled waveguides~\cite{Burr2015,Othman2016,Veysi2018}, and it leads to a faster growth of $Q$-factor as $Q\propto N^5$~\cite{Figotin2005,Zhang2020}. Interestingly, even further decrease of the period results in a change of the sign of the group velocity dispersion, as shown schematically in Fig.~\ref{fig:1}.
As a result, the dependence $Q(N)$ becomes nonmonotonic and preserves for arbitrarily large $N$ (see Fig.~\ref{fig:1}, solid curves). Three values of the period $a_1, a_2, a_3$ are chosen so that the calculated $Q(N)$ curves exhibit the first maximum for $N=19$, $N=22$, and $N=29$, respectively. However, we notice that by fine tuning of the period it is possible to shift the first maximum to arbitrary number $N$. The maximal $Q$-factor, obtained separately for each value of $N$ then becomes approximately proportional to $ N^{\sim 6.88}$, see upper dashed line in Fig.~\ref{fig:2}(a). Such a seemingly slight modification allows increasing the $Q$ factor by several orders of magnitude even for short arrays, $N \approx 10$.


{\it Spherical multipole expansion.}
One of the most powerful and universal tools used in nanophotonics for describing radiative properties of finite structures is based on the multipolar content analysis. Despite that the considered system consists of dipolar emitters each with a single multipolar channel of radiation, the overall multipolar content of the collective modes can be arbitrary complex. However, it turns out, that the observed suppression of radiation has a clear signature in the multipolar picture, similar to the quasi-bound states in the continuum resonances~\cite{Rybin2017,Bulgakov2020}, where lower-order multipole moments are suppressed.  To elaborate more on that, we perform the multipolar decomposition of the eigenmodes of the array of $N=10$ dipoles in terms of the vector spherical harmonics (VSH)  basis as described in Supplemental Materials~\cite{SuppMat}.
 The total radiative decay rate $ \Gamma$ normalized to the radiative decay rate of an individual dipole $\gamma_0$ can be expanded into a sum containing all multipole channels labeled by the subscript $j$ (dipole $j=1$, quadrupole $j=2$, octupole $j=3$, etc.):  ${\Gamma} =\gamma/\gamma_0= \sum\limits_{j=1}^{\infty} {\Gamma}_{j}$, where ${\Gamma}_{j}$ can be obtained as follows:


\begin{figure}[t]
\centering
\includegraphics[width=0.5\textwidth]{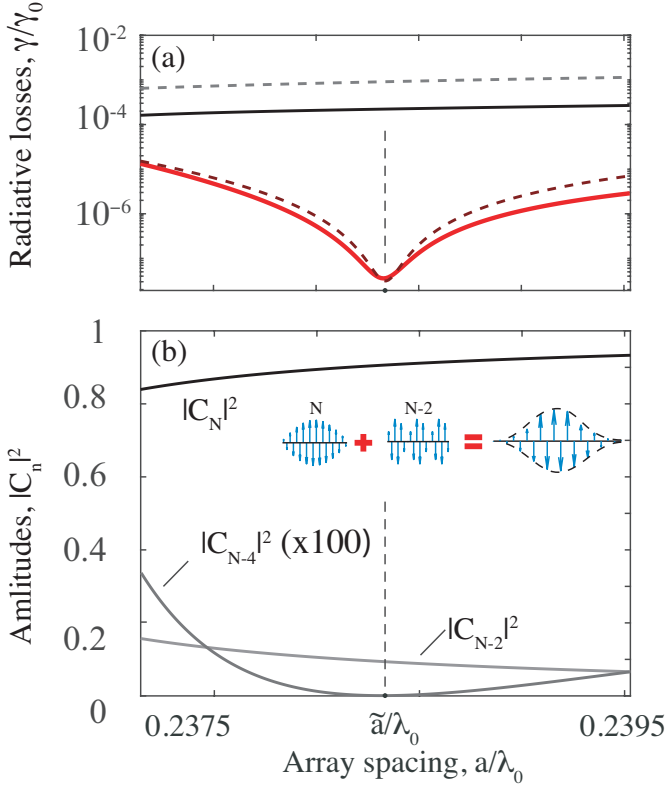}
\caption{
(a) Normalized radiative losses rate $\gamma/\gamma_0$ as function of the distance $a$ for $N$ and $N-2$. Solid lines correspond to exact full solution, while dashed lines correspond to approximate coupled mode model. (b) Decomposition amplitudes of the high-$Q$ band-edge state showing that the radiation cancellation occurs due to the interference of $N$ and $N-2$ modes onle ($C_{N-4}=0$ at $a= \tilde a$) illustrated in the inset.}
\label{fig:3}
\end{figure}

\begin{eqnarray}
 \Gamma_j = \dfrac{3}{8 \pi} \dfrac{1}{ |\mathbf{d}|^2 k_0^6 } \sum\limits_{m = - j}^{+ j} \left( |a_{j, m}|^2 + |b_{j, m}|^2 \right),
\label{gammaexpan}
\end{eqnarray}
and $a_{j, m}$ and $b_{j,m}$ are electric and magnetic multipolar amplitudes [see Supplemental Materials \cite{SuppMat}, Eq.~(S10)], $d$ is the amplitude of an individual dipole, and $k_0$ is the wavenumber corresponding to resonant frequency $\omega_0$.  The total radiative decay rate and its multipole decomposition for the band edge states are shown in Fig.~\ref{fig:2}(b) as a function of the period $a$. The reported high-$Q$ state formed for the optimal distance $\tilde a/\lambda_0\sim 0.24$ (indicated by dashed vertical line) corresponds to the strongest deep in total radiative decay rate. Surprisingly, it turns out that all the contributions of VSH with order $j$ up to approximately $N$ are also resonantly suppressed at this optimal distance. We notice that for other separation distances there are dips in the radiative emission rate provided by simultaneous suppression of smaller numbers of multipoles. The described effect has similarities with other collective effect such as superscattering~\cite{Ruan2010} and zero scattering with high-order anapole states~\cite{Valero}.

{\it Mode coupling approach.} The origin of the described cancellation of high-order multipoles lies in the destructive interference of the standing waves of a dipolar array which is attributed to their external coupling as schematically shown in Fig.~\ref{fig:1}. Such coupling can occur in open systems due to their interaction via far-field radiation. Initially proposed in quantum mechanics~\cite{FriedrichH.1985}, external mode coupling has been actively developed in nanophotonics since recently~\cite{Cao2015} for formation of high-$Q$ states in individual resonators~\cite{Wiersig2006,Rybin2017,Koshelev2020} and their ensembles~\cite{Bulgakov2020}, including bound states in the continuum that occur in infinite systems above the light cone.


In the considered finite periodic array of dipoles the eigenmodes can be represented as two counter propagating polaritonic Bloch modes formed from the hybridization of dipolar excitation and free space modes \cite{Weber2004}. The solution of the system of equations for coupled dipoles in nearest neighbour interaction approximation provides us with the distributions of the dipole moments, which form a full set of functions on a discrete array~\cite{Malyshev1995,Weber2004,Ivchenko2013}:
\begin{equation}
d^{(l)}_{k} = \sqrt{\frac{2}{N+1}} \sin \left( \dfrac{k \pi}{N+1} l \right), \quad
 k=1..N,
 \label{eq:TBA_modes}
\end{equation}
where $l$ labels the scatterer, and $k$ is the eigenstate number. In the general case of long-range interaction, however, such approximation fails, and that is the major factor that affects the formation of high-Q resonant modes.

The effective Hamiltonian describing the dipole-dipole interaction of the transverse dipolar excitations in the considered array can be written  in the following form:
\begin{equation}
\label{eq:H}
\widehat H=-\sum_{n}i\gamma_0/2 \dyad{n}{n}+\sum_{mn,\ m\neq n} g_{mn}\dyad{m}{n}+\text{h.c.}
\end{equation}
where $\ket{n}$ is the Fock state of a single excitation at the dipolar site with number  $n$, and $g_{mn} = -3 \pi \gamma_0 k_0^{-1} G_{0,tt}(\mathbf{r_m}, \ve{r_n}, \omega_0)$ is the dipole-dipole coupling coefficient defined by transverse component of the dyadic Green's function. The eigenmodes of the dipolar excitation in the array can be represented as the solution of Schrodinger equation $\widehat H \ket{\psi}= (\omega-i\Gamma)\ket{\psi}$. In order to construct them, we use the conventional nearest neighbour basis provided by the expression~\eqref{eq:TBA_modes} and expand the unknown eigenvector over the chosen basis $\ket{\psi}=\sum_k c_k \ket{\varphi_k}$, where $ \ket{\varphi_k}=\sum_{n=1}^{N} d_k^{(n)} \ket{n}$. The Schrodinger equation yields a system of equations for the coupling coefficients:
\begin{equation}
(\omega_n-i\Gamma_n)c_n=\matrixel{\varphi_n}{\widehat H}{\varphi_n} c_n+\sum_{m\neq n} \matrixel{\varphi_n}{\widehat H}{\varphi_m}c_m.
\label{eq:SchrEq}
\end{equation}

\begin{figure}[t]
\centering
	\includegraphics[width=0.48\textwidth]{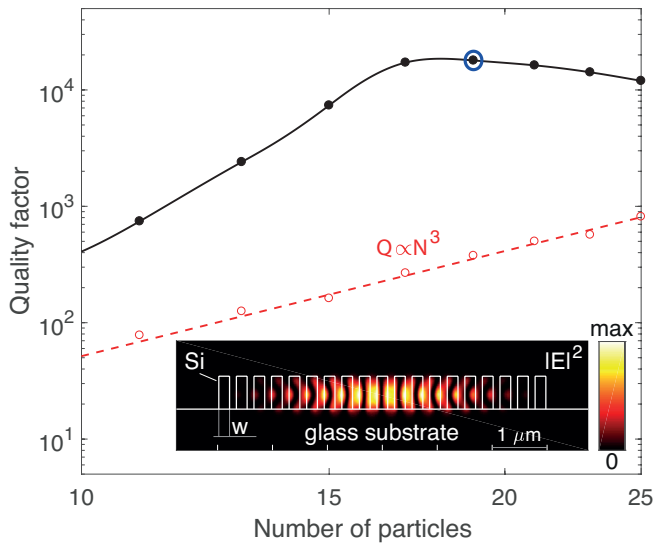}
	\caption{Dependence of the $Q$-factor of the resonance with the highest $Q$ on the number of the particles $N$ in an array of Si rectangular bars with height and width equal to 600 and 550 nm, respectively. The period of the structure is fixed to $a=320$ nm, while thickness of each element $w$ is varied. Red dashed line corresponds to band-edge states ($w=190$ nm), while solid black line and circles correspond to the high-$Q$ state ($w_{highQ}=197$ nm). Inset shows the field distribution of electric field intensity in the array of 19 particles (side view) for the maximal value of $Q$-factor marked with blue circle. Refractive index of silicon and glass were taken equal to $n_{Si}=3.5$ and $n_{glass}=1.45$. }
	\label{fig:Numerical}
\end{figure}

Now, the system~\eqref{eq:SchrEq} that describes the coupling of different standing waves in the dipolar array can be applied  for reconstructing the high-$Q$ states in terms of nearest-neighbour  solution~\eqref{eq:TBA_modes}. It turns out, that though the exact decomposition should take into account all basis modes, a simplified expansion involving  only three modes is sufficient to describe the considered high-Q state: $\ket{\psi_{highQ}}=c_N\ket{\phi_N}+c_{N-2}\ket{\phi_{N-2}}+c_{N-4}\ket{\phi_{N-4}}$.  In such an expansion, the terms $N-1, N-3$ are missing as they have different symmetry and are not coupled through the Hamiltonian \eqref{eq:H}, and matrix \eqref{eq:SchrEq} is reduced to $3\times3$ system. By solving the system numerically one can derive the radiative losses rates of the states as shown in Fig.~\ref{fig:3}(a), which now can be compared with the solution of the exact coupled dipole system [see Supplemental materials~\cite{SuppMat})]. We notice that the appearance of radiative losses repulsion typical for the external coupling scheme~\cite{Cao2015}. The relative losses rate of the subradiant mode drop down to $10^{-8}$ for the specific distance $\tilde a$. At the same time, we plot the eigenfrequencies and losses obtained from the reduced system \eqref{eq:SchrEq} with dashed and dash-dotted lines having perfect matching for real and imaginary parts of the high-$Q$ mode energies. Moreover, the expansion amplitudes $|c_N|^2, |c_{N-2}|^2, |c_{N-4}|^2$ plotted in Fig.~\ref{fig:3}(b) show that at the point of minimal losses $C_{N-4}$ vanishes, which means that the high-$Q$ state consists only of two partial modes with number $N$ and $N-2$, as depicted in Fig.~\ref{fig:1}. The spectral alignment of the corresponding modes of a infinite array results in the destructive far-field interference and suppression of the total radiation rate of the mode of a finite array. A similar effect has been observed for two-photon states in a closely spaced arrays of cold atoms placed near a waveguide~\cite{Poddubny2020} and for the split-band edge resonances in photonic crystals~\cite{NohPRA2010}.

{\it Silicon-on-insulator structure.}
So far, we have discussed an ideal model of coupled point dipoles. In order to show the application of the described effect in realistic dielectric optical resonators, we design a nanostructure with maximized $Q$ factor that can be readily fabricated from the silicon-on-insulator wafers by means of common fabrication techniques (see inset in Fig.~\ref{fig:Numerical}).
All geometric parameters of the structure are fixed except for the thickness $w$, which is employed for the fine tuning of the $Q$-factor. We calculate the $Q$ factors of the magnetic dipole eigenmodes of the finite arrays with varying $N$ and two values of the thickness $w$. Results presented in Fig.~\ref{fig:Numerical} show that for a specific value $w = w_{highQ}$, the $Q$ factor grows much faster than that of a regular band-edge state (observed for $w=w_{BE}$), and it reaches a maximum at $N=19$, similarly to the case of Fig.~\ref{fig:2}(a). The $Q$-factor of the high-$Q$ state exceeds that in the band-edge regime by almost two orders of magnitude reaching the value $\approx 2\cdot10^4$. Due to interference of two standing waves, the field distribution of the mode becomes more localized as shown in the inset of Fig.~\ref{fig:Numerical}. As a result, the mode volume becomes as small as $\sim 2.5 (\lambda/n)^3$, implying the Purcell factor $\sim 560$, which is much larger than the values reported previously for similar structures~\cite{Krasnok2016,Hoang2020}.  Moreover, by tuning the parameters of the array one can get the optimal conditions for $N=29$ with higher $Q$-factor of $Q=1.8\cdot 10^{5}$ (see Supplemental Materials \cite{SuppMat}) and the Purcell enhancement factor of $\sim 3400$. Note, that the structure has subwavelength dimensions in two transverse directions and only a few wavelengths in the longitudinal dimension.

Importantly, this new physical mechanism of the $Q$-factor enhancement can be compared with interference of two leaky modes supported by isolated dielectric resonators~\cite{Koshelev2020}, and also with
destructive interference of the dipole and toroidal modes and the formation of the so-called {\it optical anapoles}~\cite{anapole}. In addition, we should mention some similarities with the physics of the so-called {\it embedded solitons} in nonlinear modes with higher-order dispersion~\cite{soliton,soliton_book}.

In summary, we have revealed a new mechanism of the formation of high-$Q$ states in finite-extend subwavelength photonic systems supporting non-monotonic dispersions. The interference of the band-edge mode and another standing mode of the finite periodic nanoparticle array leads to the cancellation of the far-field radiation through simultaneous suppression of multipole components up to the order $N$, where $N$ is the number of the elements in the array. We have demonstrated
how this general concept can be realized with an array of Mie-resonant silicon nanoparticles with dramatic suppression
of radiation losses and a giant enhancement of the Purcell factor by tuning of the array parameters. We believe that our finding will extend the existing approaches in a design of high-$Q$ photonic resonant subwavelength structures.

\begin{acknowledgments}
The authors acknowledge a financial support from the Russian Foundation of Basic Research (grant 20-52-12062), the Australian Research Council (grant DP210101292), the Strategic Grant of the Australian National University. They also thank A. Bogdanov, P. Belov, and A. Sheremet for useful discussions.
\end{acknowledgments}

\end{document}


\title{High-Q localized states in finite arrays of subwavelength resonators \\Supplemental Material}

\author{Danil~F.~Kornovan}
\affiliation{\affilITMO}

\author{Roman~S.~Savelev}
\affiliation{\affilITMO}

\author{Yuri~S.~Kivshar}
\affiliation{\affilITMO}
\affiliation{\affilANU}

\author{Mihail~I.~Petrov}
\affiliation{\affilITMO}


\newcommand{\affilITMO}{Department of Physics and Engineering, ITMO University,  St.-Petersburg 197101, Russia}
\newcommand{\affilANU}{Nonlinear Physics Center, Research School of Physics, Australian National University, Canberra ACT 2601, Australia}

\maketitle
\tableofcontents

\section{I. Dipole model}
\label{app:model}
The self-consistent system of equations describing the interaction of oscillating dipole moments in a one-dimensional periodic array can be expressed as~\cite{ Weber2004, Petrov2015}:
\begin{eqnarray}
\mathbf{d}^{(n)} = 4 \pi k^2 \alpha(\omega)  \sum\limits_{l = 1, l \ne n}^{N} \mathbf{G}(\mathbf{r}_n, \mathbf{r}_l, \omega) \mathbf{d}^{(l)},
\label{eq:CDE}
\end{eqnarray}
where $\mathbf{d}^{(n)}$ is the dipole moment of the $n$-th particle, super-/subscripts $n, l$ enumerate the dipoles, $k$ is the wavenumber, $\mathbf{G}( \mathbf{r}_n, \mathbf{r}_l, \omega)$ is the classical electromagnetic Green's tensor \cite{Novotny2012}. In the vicinity of the resonance, the polarizability of the resonator can be approximated with~\cite{LoudonJoPBAMOP2006}:
\begin{eqnarray}
\alpha(\omega) = - \alpha_0 \dfrac{\Gamma_0/2}{\Delta + i \Gamma_0/2},
\label{atompol}
\end{eqnarray}
where $\Gamma_0$ is the decay rate of the individual resonator, and $\Delta = \omega - \omega_0$ is the detuning between the external excitation and dipole resonance frequencies. Assuming that $\omega_0 \gg \Gamma_0$, we can apply quasiresonant approximation setting that  i) $\mathbf{G}( \mathbf{r_n}, \mathbf{r_l}, \omega)\approx \mathbf{G}( \mathbf{r_n}, \mathbf{r_l}, \omega_0)$, and ii) $\alpha_0 \approx \frac{2}{3} k_0^{-3}$, with $k_0 = \omega_0/c$. Within this assumption the system~\eqref{eq:CDE} is linearized to a regular eigenvalue problem, which solution provides one with the complex eigenfrequencies $\omega=\omega' + i\Gamma$ and the distribution of the dipole moments of the resonant states supported by a dipolar array. Since the real part of the complex eigenfrequency $\omega'$ weakly differs from the resonance frequency of an individual dipole $\omega_0$, the normalized quality factor of an eigenmode can be approximated as $Q/Q_0 \approx \Gamma_0/\Gamma$.

\section{II. Dispersion of an infinite array}
\label{app:dispersion}
The dispersion of eigenmodes in an infinite periodic chain of dipoles is described by a solution of the well-known analytical equation~\cite{Citrin2006, Asenjo2017}: 
\begin{multline}
\dfrac{\Delta\omega}{\Gamma_0} =  \dfrac{3}{4 k_0^3 a^3} \text{Re}\: 
\bigg[ \text{Li}_{3}( e^{i(k_0 + q)a} ) + \text{Li}_{3}(e^{i(k_0 - q)a}) - \\  i k_0 a\left[ \text{Li}_2(e^{i (k_0 + q) a} + \text{Li}_2(e^{i (k_0 - q) a})  \right] - \\
  (k_0 a)^2 \left[ \text{Li}_1(e^{i (k_0 + q) a} + \text{Li}_1(e^{i (k_0 - q) a} \right] \bigg],
\label{ICHD} 
\end{multline}
where  $q$ is the eigenmode wavenumber, and Li$_n(x)$ is the polylogarithm of order $n$.

One can take a closer look at the behaviour of the dispersion relation in the vicinity of the band edge $\pi/a$, where it contains only even powers of $\Delta q = q - \pi/a$:
\begin{align}
\dfrac{\Delta\omega(\Delta q \approx 0)}{\Gamma_0} = c_2 (k_0 a) \cdot (\Delta q a)^2 + c_4 (k_0 a) \cdot (\Delta q a) ^4 + ..., \nonumber\\
c_2 (k_0 a) = 4\text{Ln}\left[2 \left|\cos\left( \frac{k_0 a}{2} \right) \right| \right] + \dfrac{k_0 a \left( \sin\left(k_0 a\right) - k_0 a \right) }{\cos^2 \left( \dfrac{k_0 a}{2} \right)}.
\label{c2eq}
\end{align}
The coefficient $c_2$ as a function of $k_0 a$ has a single zero at $\dfrac{k_0 a}{2 \pi} \approx 0.2414...$, which is the limiting value of the period allowing to observe strong suppression of the emission in a chain of $N$ atoms for $N \to \infty$, as was  reported previously \cite{Kornovan2019}. Therefore, the dispersion curve for the period defined by Eq.~\eqref{c2eq} at the band edge behaves as $\Delta \omega \sim (q - \pi)^4$. 

\section{III. Multipole expansion}
\label{app:mltpls}

Solution of the system \eqref{eq:CDE} yields the distributions of dipole moments in the eigenmodes $\mathbf{d^{(l)}}$. The radiated field at the point $\mathbf{r}$, which does not coincide with the dipoles positions $\mathbf{r_l}$,  can be represented as:
\begin{eqnarray}
\label{eq:app_mlt_exp}
\mathbf{E}(\mathbf{r}, \omega_0) = 4 \pi k_0^2 \sum\limits_{l = 1}^{N} \mathbf{G_0}(\mathbf{r}, \mathbf{r_l}, \omega_0) \mathbf{d^{(l)}}.
\end{eqnarray}

In order to make the spherical multipole expansion of the field $\mathbf{E}(\mathbf{r}, \omega_0)$, we substitute  the expression of the Green's tensor written in terms of vector spherical harmonics (VSH) \cite{Chew1995}:
\begin{eqnarray}
& \mathbf{M}_{j, m}(\mathbf{r}, k) = \nabla \times \mathbf{r} j_j(k r) Y_{j, m}(\theta, \phi), \quad \mathbf{N}_{j, m}(\mathbf{r}, k) = \dfrac{1}{k} \nabla \times \mathbf{M}_{j, m}(\mathbf{r}, k), \nonumber\\
& \mathbf{G_{0}}(\mathbf{r}, \mathbf{r'}, \omega) = - \dfrac{\mathbf{e_{r}} \otimes \mathbf{e_{r}}}{k^2} \delta(\mathbf{r} - \mathbf{r'}) + i k \sum\limits_{j = 1}^{\infty} \sum\limits_{m = - j}^{+ j} \dfrac{1}{j (j + 1)} \times \nonumber\\
&   \begin{cases}
\mathbf{M^{(1)}_{j, m}}(k, \mathbf{r}) \otimes \mathbf{M_{j, -m}}(k, \mathbf{r'}) + \mathbf{N^{(1)}_{j, m}}(k, \mathbf{r}) \otimes \mathbf{N_{j, -m}}(k, \mathbf{r'}), \quad \text{if $r > r'$} \\
\mathbf{M_{j, m}}(k, \mathbf{r}) \otimes \mathbf{M^{(1)}_{j, -m}}(k, \mathbf{r'}) + \mathbf{N_{j, m}}(k, \mathbf{r}) \otimes \mathbf{N^{(1)}_{j, -m}}(k, \mathbf{r'}), \quad \text{if $r < r'$}
\end{cases}
\end{eqnarray}
where $j_j(x)$ are the spherical Bessel functions of order $j$, $Y_{j,m}(\theta, \phi)$ are the spherical harmonics, and superscript $^{(1)}$ denotes that spherical Bessel function $j_{j}(kr)$ has to be replaced with the spherical Hankel function of the first kind $h^{(1)}(kr)$.

The Green's function tensor  $\mathbf{G_0}(\mathbf{r}, \mathbf{r'}, \omega)$ expansion through VSH depends upon how the field point $\mathbf{r}$ is located with respect to $\mathbf{r'}$. Our goal is to compute the total power radiated into the far-field, so, we are interested in the field at point $r \gg r_l$ for any $l$, therefore, in what follows we fix the form of the expansion. Then we plug the Green's tensor expansion into \eqref{eq:app_mlt_exp} and obtain:
\begin{align}
\mathbf{E} (\mathbf{r}, \omega) = \sum\limits_{j = 1}^{\infty} \sum\limits_{m = - j}^{+ j} \sum\limits_{l = 1}^{N} \left( a_{j, m}^{(l)} \dfrac{i \mathbf{N^{(1)}_{j, m}}(k, \mathbf{r})}{\sqrt{j (j + 1)}}  + b_{j, m}^{(l)} \dfrac{i \mathbf{M^{(1)}_{j, m}}(k, \mathbf{r})}{\sqrt{j (j + 1)}} \right),
\end{align}
where $a_{j, m}^{(l)}, b_{j, m}^{(l)}$ are the complex-valued scalars being proportional to the scalar products of right dyads entering the Green's tensor and $l^{\text{th}}$ paricle dipole moment, which have the following explicit forms:
\begin{eqnarray}
a_{j, m}^{(l)} = \dfrac{4 \pi k_0^3}{\sqrt{j(j+1)}} \mathbf{N_{j, -m}}(k_0, \mathbf{r_l}) \cdot \mathbf{d^{(l)}}, \nonumber\\
b_{j, m}^{(l)} = \dfrac{4 \pi k_0^3}{\sqrt{j(j+1)}} \mathbf{M_{j, -m}}(k_0, \mathbf{r_l}) \cdot \mathbf{d^{(l)}}.
\end{eqnarray}

After that, we calculate the period-averaged Poynting vector and the total normalized radiated power:
\begin{eqnarray}
& \langle \mathbf{S} \rangle = \dfrac{c}{8 \pi} \text{Re}\left[ \mathbf{E} (\mathbf{r}, \omega) \times \mathbf{H^{*}} (\mathbf{r}, \omega)  \right], \nonumber\\
& {\Gamma} = \dfrac{P}{P_1} = \dfrac{ \lim\limits_{r \to \infty} \int \int \langle \mathbf{S} \rangle \cdot  \mathbf{n_r} \: r^2 \: d \Omega }{\lim\limits_{r \to \infty} \int \int \langle \mathbf{S_{1}} \rangle \cdot  \mathbf{n_r} \: r^2 \: d \Omega },
\end{eqnarray}
with $P_1 = c |\mathbf{d}|^2 k_0^4 / 3$ corresponding to a power irradiated by a single dipole $\mathbf{d}$ ($c$ is the speed of light), integration is done over a spherical surface of radius $r$, and $\mathbf{n_r}$ being normal to it.

After some algebraic manipulations  and taking into account the orthogonality of the VSH functions, it can be shown that this quantity can be simply expressed as:
\begin{eqnarray}
{\Gamma} = \dfrac{3}{8 \pi} \dfrac{1}{ |\mathbf{d}|^2 k_0^6 } \sum\limits_{j = 1}^{\infty} \sum\limits_{m = - j}^{+ j} \left( |a_{j, m}|^2 + |b_{j, m}|^2 \right),
\label{gammaexpan}
\end{eqnarray}
with $a_{j, m} = \sum\limits_{l=1}^{N} a_{j, m}^{(l)}$, $b_{j, m} = \sum\limits_{l=1}^{N} b_{j, m}^{(l)}$. Essentially, \eqref{gammaexpan} allows one to observe how a VSH with $(j,m)$ contributes to the total emission rate. Since the total number of harmonics grows quadratically with increasing $j$, we would like to rather look at contributions of harmonics with a given $j$: $\Gamma = \sum\limits_{j=1}^{\infty} \Gamma_{j}$.

One important note on the multipole expansion has to be made: the expansion coefficients $a_{j, m}, b_{j, m}$ depend upon how the system is oriented with respect to the introduced coordinate system. Of course, it does not affect the observable quantities such as ${\Gamma}$, but alters  ${\Gamma}_j$, and, therefore, it affects the convergence of ${\Gamma}$ with $j$.


\section{IV. Finite extent arrays of dielectric resonators}

\begin{figure}[t]
\begin{center}
	\includegraphics[width=1.0\textwidth]{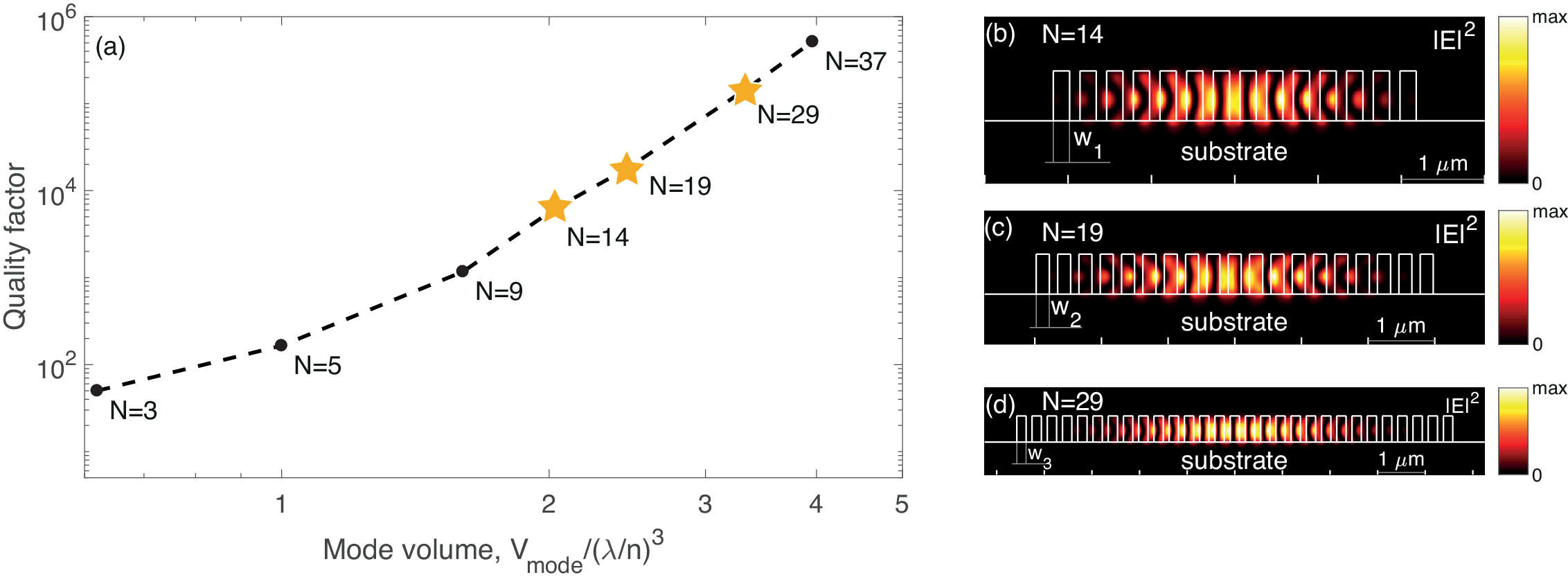}
	\caption{ High-$Q$ arrays of Si nanoparticles. (a) Maximum values of the $Q$-factor and corresponding normalized mode volumes calculated for different number of particles $N$ for one-dimensional arrays of Si nanoparticles placed on a glass substrate. Parameters of the structure are given in the caption to Fig.~4 of the main text; $w$ is varied from 252\;nm (maximum $Q$ for $N=3)$ to 195\;nm (maximum $Q$ for $N=37)$. (b-d) Electric field intensity  for the localized states with the highest values of $Q$ for $N=14$, $N=19$, and $N=29$, respectively. These states are marked in (a) by orange stars.}
	\label{fig:QvsV}
\end{center}

\end{figure}

Figure~4 of the main text shows the dependence of the quality factor $Q$ on the number of particles $N$ for a particular value of the thickness of the particles $w$, when the maximum of $Q$ is achieved for $N=29$.  In Fig.~\ref{fig:QvsV}(a), we show the maximum $Q$ factors for arrays with finite extension varying from $N=3$ to $N=37$, as well as the corresponding mode volumes. The mode volume is calculated as follows:
\begin{equation*}
    V=\dfrac{\int \eps(\VB{r})\VB{E}(\VB{r})^2dV}{\mathrm{max}(\eps(\VB{r})\VB{E}(\VB{r})^2)},
\end{equation*}
where normalization is performed over the whole calculation domain, and in Fig.~\ref{fig:QvsV} the result is further normalized by the third power of reduced wavelength $\lambda/n$, where $n$ is the refractive index of silicon. The examples of electric field intensity distributions are shown in Figs.~\ref{fig:QvsV}(b-d) for the points $N=14$, $N=19$, and $N=29$ in (a), respectively. The corresponding values of $w$ are 199~nm, 197~nm, and 195.5~nm. Importantly, slower than linear growth of the mode volume results in a dramatic enhancement of the Purcell  factor from moderate values of $\lesssim 10$ for the short arrays with $N \lesssim 10$ particles to $10^4$ for arrays with $N\approx 40$ particles.

